\documentclass[3p,times,procedia]{elsarticle}
\usepackage[bookmarks=false]{hyperref}
    \hypersetup{colorlinks,
      linkcolor=blue,
      citecolor=blue,
      urlcolor=blue}
\usepackage{tikz}
\usetikzlibrary{positioning,arrows.meta,fit,backgrounds,calc,shapes.geometric}
\usepackage{fontawesome5}
\usepackage{booktabs}

\usepackage{amssymb}

\usepackage[figuresright]{rotating}

\begin{document}
\begin{frontmatter}




\title{A Modular Workflow for Multimodal Reading Experiments}


\author[a]{Thomas Krämer\corref{cor1}} 
\author[b]{Thomas Kosch}
\author[a]{Dagmar Kern}
\author[a]{Daniel Hienert}

\address[a]{GESIS - Leibniz Institute for the Social Sciences, Unter Sachsenhausen 6-8, 50667 Cologne, Germany}
\address[b]{HU Berlin, Unter den Linden 6, 10099 Berlin, Germany}

\begin{abstract}
We introduce a web-based modular workflow for real-time multimodal experiments in naturalistic online reading. The workflow integrates eye tracking, EEG, and interaction data from mouse and keyboard, synchronizes them via Lab Streaming Layer, and links gaze to browser-based text at the word, sentence, and AOI levels. It is designed as a reusable experimental procedure that can be adapted to different sensors, tasks, and analysis goals. As a use case, we apply the workflow to a study of selective exposure in online news search and reading. During the experiment, gaze-derived measures are computed online, while EEG and other synchronized streams are processed immediately after task sessions based on fixation-triggered segmentation. The resulting behavioral, neural, and linguistic metrics support selecting text passages for targeted post-task rating or labelling within the same lab session. The workflow thus provides a general basis for multimodal research on reading and related cognitive processes, and supports the empirical validation, in ecological contexts, of constructs that are typically operationalised through self-report measures.
\end{abstract}

\begin{keyword}
Selective Exposure; Electroencephalography; Eye Tracking; Co-Registration; News; Naturalistic Reading




\end{keyword}
\cortext[cor1]{Corresponding author}
\end{frontmatter}

\email{thomas.kraemer@gesis.org}



\section{Introduction}

Studying naturalistic reading with different input modalities in a web-based context remains challenging. A first key difficulty concerns the constructs themselves: many rely on self-report measures, which are susceptible to bias and whose validity in ecological reading contexts remains an open question. A second challenge lies in analysing and processing co-registered, multimodal data in real time so that the resulting insights can be incorporated directly into the ongoing study
In this paper, we present an extensible workflow that enables fine-grained multimodal analysis of naturalistic online reading by jointly capturing behavioural, physiological, and interaction data, while making these insights immediately available for use within the experimental procedure.
In our implementation, we combine eye-tracking, electroencephalography (EEG), and interaction data as multimodal input streams. Eye tracking captures reading behaviour in real time, while EEG provides neural metrics of information processing, for example, through frequency-based measures or fixation-related potentials (FRPs). The framework dynamically aligns eye-tracking data with the textual stimulus at the word and sentence level, enabling analyses, for instance, of how eye-tracking and EEG signals correlate or diverge. It also allows researchers to define criteria for selecting sentences for further investigation. Additionally, textual stimuli can be automatically annotated with linguistic features during processing, providing partial control over stimuli that are not determined in advance in naturalistic reading scenarios. 

So far, we have applied this pipeline to investigate cognitive biases, particularly selective exposure. However, we see broader potential for studying naturalistic reading in web environments, including applications such as emotion detection, identification of comprehension difficulties, support for foreign language learning, and examination of perceptions of different writing styles.

By combining multimodal co-registration, linguistic characterisation of textual stimuli, and metric-based re-presentation of relevant passages within a single session, we provide an instrument for reading experiments that can use EEG and eye-tracking as proxies for underlying cognitive constructs, enabling testing of the validity of these signal-construct relationships in naturalistic settings. 
The pipeline links recorded multimodal signals to linguistically annotated text stimuli at the sentence level, enabling detailed post-hoc analyses that connect physiological responses to textual characteristics and providing a versatile foundation for studying online reading under ecologically valid conditions.

\section{Related work}
Multimodal sensing approaches that combine physiological and behavioural signals have increasingly been used to investigate cognitive and affective states during media consumption and human–computer interaction. In particular, co-registration of electroencephalography (EEG) and eye tracking is considered the only viable approach for studying EEG correlates in naturalistic reading scenarios, given the complex nature of reading behaviour\cite{Dimigen2011coregistration}. In this section, we review prior work that combines EEG and gaze data for emotion detection, confusion detection during reading, and the investigation of cognitive biases in HCI contexts.

\subsection{Multimodal EEG and Eye Tracking for Emotion Detection}
In the past decade, emotion detection based on multimodal sensor data has received growing attention.
Recent review work~\cite{Pillalamarri2025,JAFARI2023107450} summarizes studies and datasets that combine EEG with other modalities for emotion recognition. 
In the summary by Jafari et al.~\cite{JAFARI2023107450}, only two datasets include eye-tracking data (SEED-IV~\cite{Zheng2018Emotionmeter} and MAHNOB-HCI~\cite{soleymani2011multimodal}). Both datasets use video stimuli trying to classify viewers' emotions based on EEG and eye movements. Similarly, the review by Pillalamarri and Shanmugham~\cite{Pillalamarri2025} identifies only two out of nine studies~\cite{Zhao2019,Su2019} that employ gaze as one of the modalities\footnote{Only two out of seven datasets (SEED~\cite{zheng2015investigating,duan2013differential}, and SEED-IV) include gaze information}. Zhao et al.~\cite{Zhao2019} investigated the complementary representation properties of EEG and eye movements recorded while participants watched videos with highly emotional content. They classified five emotional states (happy, sad, fear, disgust, and neutral) and found that combining features using a bimodal deep auto-encoder improved classification accuracy (79.7\%) compared to using EEG alone ($\approx-10\%$) or eye movements alone ($\approx-20\%$). 

\subsection{Multimodal EEG and Eye Tracking for Detecting Confusion}
Beyond emotion detection, multimodal physiological sensing has also been applied to identify cognitive states such as confusion during reading. Zhuang et al.~\cite{ZhuangMaes2025detectingreadinginducedconfusionusing} investigated the detection of reading-induced confusion, with predefined trials and co-registering EEG and eye-tracking data. Participants read paragraphs that were either correct and easy to comprehend, or designed to either evoke factual confusion (e.g., non-sense, contradicting common knowledge) or contextual confusion (i.e., correct but difficult to understand without substantial prior knowledge). They trained different classifiers using EEG features, gaze features, and their combination, employing XGBoost or Convolutional Neural Network (CNN) for a binary classification of confusion state. Their results show that EEG and gaze data provide complementary information, achieving an average classification accuracy of 77.3\%.

\subsection{Cognitive Biases in HCI}
Similarly, research on cognitive biases has received increasing attention in recent years, particularly amid the growing societal influence of digital media. In their systematic review of empirical research on cognitive biases in Human-Computer Interaction (HCI), \cite{Boonprakong2025a} identified several application contexts, including information interaction and recommender systems, human-AI interaction, and behaviour change. The reviewed studies focus on investigating biases, observing their effects, mitigating them, quantifying them, or utilizing them in system design. However, only 2 of 127 reviewed studies employ sensor-based measurements at all. 

\subsection{Research Gap}
Given the consistently stronger classification performance of multimodal approaches, it is notable that the co-registration of eye tracking with other sensor modalities, especially EEG, remains under-explored. This gap is particularly striking in studies for natural reading of self-selected texts, where eye tracking is essential for linking specific textual input to moment-by-moment behavioral and neural responses, and thus for interpreting effects, such as changes in attitudes or behavior.

However, to our knowledge, no existing approach integrates the co-registration of EEG, gaze, and other sensor data with a real-time analysis pipeline for naturalistic online reading that simultaneously links gaze to the browser DOM and derives linguistic characteristics of the read text. Such integration allows sensor responses to be precisely related to textual input while controlling for stimulus heterogeneity in naturalistic reading studies.


\section{Modular Workflow for Multimodal Experiments in Naturalistic Reading}
To address this research gap, we developed a generic workflow for real-time multimodal experiments that supports self-selected reading while integrating synchronized sensor co-registration, including textual stimulus mapping and metrics. To make the workflow transferable to other multimodal studies, we describe it at two complementary levels. We first outline the overall experimental procedure and then present the underlying recording and processing pipeline. Together, these components demonstrate a generic, extensible workflow for a real-time multimodal experiment.
\subsection{Procedure}
In this section, we present the general characteristics of our workflow as a template for other experimental settings that combine multimodal input streams, namely eye-tracking data, EEG, and interaction data. 
The entire experimental workflow is implemented in a web application used for participant settings, pre- and post-experiment questionnaires, and the task sessions. All steps in the workflow can easily be adapted. 

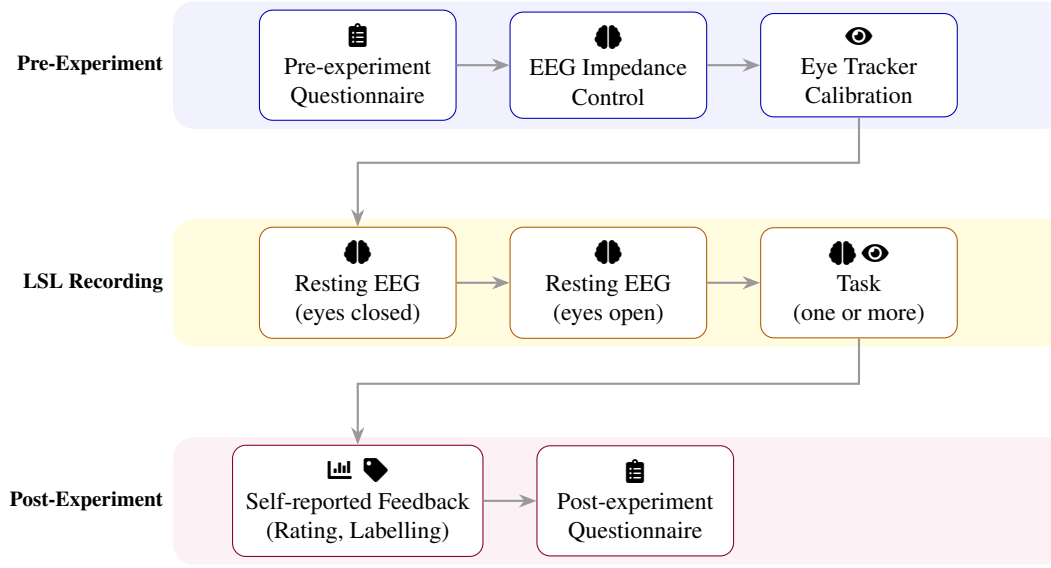
\begin{figure}[ht]
  \centering
    \begin{minipage}{1.\textwidth} 
    \begin{tikzpicture}[
        node distance=0.5cm and 0.7cm,
        base/.style={draw, rounded corners=5pt, align=center, font=\small,
                     minimum height=1.1cm, minimum width=2.6cm, inner sep=5pt, fill=white},
        pre_node/.style={base, draw=blue!70!black},
        rec_node/.style={base, draw=orange!70!black},
        post_node/.style={base, draw=purple!70!black},
        arrow/.style={-{Stealth[scale=1.0]}, line width=0.8pt, draw=gray!80}
    ]
        \coordinate (left_align) at (-1.6, 0);
        \node[inner sep=0, minimum width=0] (right_marker) at (8.5, 0) {}; 

        \node[pre_node] (p1) at (0,0) {\faClipboardList\\ Pre-experiment\\ Questionnaire};
        \node[pre_node, right=of p1] (p2) {\faBrain\\ EEG Impedance\\ Control};
        \node[pre_node, right=of p2] (p3) {\faEye\\ Eye Tracker\\ Calibration};

        \node[rec_node, below=1.4cm of p1] (r1) {\faBrain\\ Resting EEG\\ (eyes closed)};
        \node[rec_node, right=of r1] (r2) {\faBrain\\ Resting EEG\\ (eyes open)};
        \node[rec_node, right=of r2] (r3) {\faBrain \ \faEye \\ Task \\ (one or more) };

        \node[post_node, below=1.4cm of r1] (po1) {\faChartBar~~\faTag \\ Self-reported Feedback\\ (Rating, Labelling)};
        \node[post_node, right=of po1] (po2) {\faClipboardList\\ Post-experiment\\ Questionnaire};

        \draw[arrow] (p1) -- (p2); \draw[arrow] (p2) -- (p3);
        \draw[arrow] (r1) -- (r2); \draw[arrow] (r2) -- (r3);
        \draw[arrow] (po1) -- (po2);

        \draw[arrow] (p3.south) -- ++(0,-0.6) -| (r1.north);
        \draw[arrow] (r3.south) -- ++(0,-0.6) -| (po1.north);

        \begin{scope}[on background layer]
            \node[fill=blue!5, rounded corners=8pt, fit=(left_align |- p1) (right_marker |- p1),
                  inner sep=24pt, label={[black, font=\bfseries\footnotesize]left:Pre-Experiment}] (bg1) {};
            \node[fill=yellow!15, rounded corners=8pt, fit=(left_align |- r1) (right_marker |- r1),
                  inner sep=24pt, label={[black, font=\bfseries\footnotesize]left:LSL Recording}] (bg2) {};
            \node[fill=purple!5, rounded corners=8pt, fit=(left_align |- po1) (right_marker |- po1),
                  inner sep=24pt, label={[black, font=\bfseries\footnotesize]left:Post-Experiment}] (bg3) {};
        \end{scope}
    \end{tikzpicture}%
    \end{minipage}
  \caption{Experimental procedure consisting of three phases: (1) Pre-experiment, including a questionnaire, EEG impedance check, and eye-tracker calibration (2) LSL recording, involving resting EEG (eyes open/closed) and one or more tasks (3) Post-experiment, where participants give feedback on selected AOI or sentences that are selected based on metrics derived from data recorded during the task. Feedback can be made e.g. via rating or labelling the AOI, or any other feedback that is of interest. Data collection ends with a final post-experiment questionnaire.  }
    \label{fig:expflow}
\end{figure}

The experiment generally consists of three sequential phases, as illustrated in Figure \ref{fig:expflow}. The Pre-experiment phase includes completing a pre-experiment questionnaire, EEG impedance control to ensure signal quality, and calibrating the eye tracker for accurate gaze recording. The results of the pre-experiment questionnaire are stored in a database and can later be used to dynamically assign participants to different experimental conditions.

The experimental tasks are performed during the LSL Recording. Lab Streaming Layer (LSL)~\cite{Kothe2024} is a protocol and set of libraries used for temporal alignment, mapping all concurrent input streams to a common time reference. It comprises several data collection steps: an optional resting-state EEG with eyes open or closed, and one or more task sessions, e.g., reading text on different topics.

In the last post-experiment phase, participants give feedback on selected stimuli that were seen during one of the task sessions. These self-reports can be ratings of specific sentences or labelling AOIs. This feedback step is followed by a post-experiment questionnaire. It is important to note that the system processes the data collected during the LSL recording phase immediately, which is used to select sentences or AOIs on which users are requested to self-report. For this purpose, the system computes metrics used to determine a subset of viewed AOIs or sentences on which user feedback should be collected as a baseline, e.g., for later comparison of classifier performance against human judgement. 

\subsection{Recording and Data Pipeline}
\begin{figure}[htp]
  \centering
    \resizebox{0.5\columnwidth}{!}{%
      \begin{tikzpicture}[
      font=\sffamily,
      base/.style={draw, rounded corners=4pt, align=center, font=\fontsize{7pt}{8pt}\selectfont, fill=white,  minimum width=1.6cm},
      device/.style={base, draw=gray!80, fill=gray!5},
      system/.style={base, draw=gray!80},
      store/.style={base, draw=gray!80}, 
      dataset/.style={base, fill=yellow!15},
      datasetSt/.style={dataset, fill=blue!15},
      arrow/.style={-{Stealth[scale=1.0]}, draw=gray!60, line width=0.8pt},
      labelNode/.style={font=\scriptsize, fill=white, inner sep=1.5pt, text=gray!90}
    ]

      \node[device] (tobii)    at (0,0)    {\faEye\\Eye Tracker};
      \node[device] (eeggui)   at (2,0)    {\faBrain\\EEG};
      \node[device] (mouse)    at (4,0)  {\faMouse\\Mouse};
      \node[device] (keyboard) at (6,0)   {\faKeyboard\\Keyboard};

      \node[system] (lslr) at (3,-1.5) {\faNetworkWired~\faWifi \\LSL Lab Recorder};

      \node[store] (xdf) at (6,-3) {\faFileArchive \\xdf };
      \node[system] (elm) at (3,-3) {\faChrome~\faFirefoxBrowser\\EyeLiveMetrics};

      \node[store] (eegPrepoc) at (6,-4.5) {\faCogs\\Preprocesing};
      \node[store] (eegEpochs) at (6,-5.5) {\faDatabase\\EEG Segments};
      \node[store] (eegAnalysis) at (6,-6.5) {\faCogs\\EEG Analysis};
      \node[system] (elfen) at (0,-4.5) {\faCogs\\elfen};

      \node[dataset] (eegf) at (6,-7.5) {\faCreativeCommonsSamplingPlus \\EEG Metrics};
      \node[dataset] (etf) at (3,-7.5) {\faEye\\Gaze Metrics};
      \node[datasetSt] (stimMetrics) at (0,-7.5) {\faQuoteLeft\\Linguistic Metrics};
      \node[dataset] (selfreportMetrics) at (6,-9.5) {\faComment~\faPoll \\Self Reports};

      \node[system] (echoes) at (3,-9.5) {\faTag~\faTasks\\AOI / Sentence \\ Annotation / Rating};

      \draw[arrow] (tobii) -- (lslr);
      \draw[arrow] (eeggui) -- (lslr);
      \draw[arrow] (mouse) -- (lslr);
      \draw[arrow] (keyboard) -- (lslr);

      \draw[arrow] (lslr) -- (xdf) node[midway, labelNode] {all data};
      \draw[arrow] (lslr) -- (elm) node[midway, labelNode] {gaze data};

      \draw[arrow] (elm) -- (eegPrepoc) node[midway, labelNode] {word fixation onsets};
      \draw[arrow] (eegPrepoc) -- (eegEpochs);
      \draw[arrow] (eegEpochs) -- (eegAnalysis);
      \draw[arrow] (eegAnalysis) -- (eegf);
      \draw[arrow] (elm) -- (etf) node[midway, labelNode, yshift=-30pt] {per word / sentence / AOI};
      \draw[arrow] (elm) -- (elfen) node[midway, labelNode] {textual stimulus};
      
      \draw[arrow] (elfen) -- (stimMetrics);
      
      \draw[arrow] (xdf) -- (eegPrepoc) node[midway, labelNode] {raw EEG};

      \draw[arrow] (eegf) -- (echoes);
      \draw[arrow] (etf) -- (echoes);
      \draw[arrow] (stimMetrics) -- (echoes);
      \draw[arrow] (echoes) -- (selfreportMetrics);

    \end{tikzpicture}
  }
   \caption{
    Abstract overview of the multimodal experimental pipeline in a single lab session. Eye tracking, EEG, and other input streams are recorded and temporally synchronized via Lab Streaming Layer and stored in a unified data container. The EyeLiveMetrics Plugin computes gaze-based metrics and provides the fixated text, which in turn, is qualified using the elfen package. Event markers (e.g., fixation onsets) enable segmentation of neural data for time–frequency and fixation-related analyses. All derived metrics can be used for task-related annotations (e.g., on selected AOI or sentences). For example, sentences can be selected for feedback, rating, or annotation based on the longest fixation duration, or the highest increase in theta power or fixation-related potentials (FRP). }
    \label{fig:dataflow}
\end{figure}
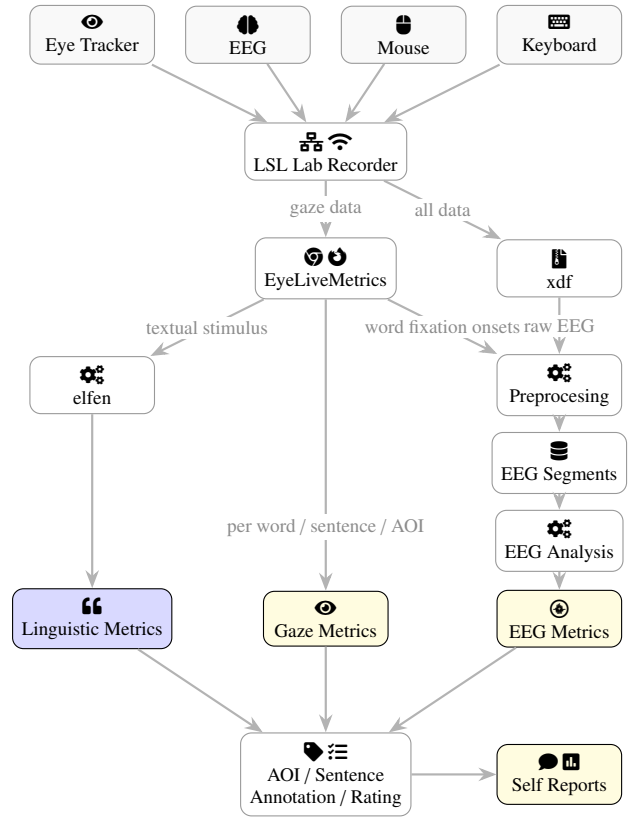

Figure \ref{fig:dataflow} provides an overview of the data collection process, during the LSL recording and post-experiment. It illustrates the synchronized multimodal recording setup, the preprocessing steps for segmenting the EEG data based on word fixation onsets, and the computation of gaze, EEG, and stimulus metrics. At the sensor level, additional LSL-compatible equipment can be easily integrated. 
Any metrics that can be derived from these data streams can be used for the post-experiment feedback: Previously viewed AOIs or sentences are first selected based on measures such as fixation duration or EEG band-power changes, and subsequently re-presented to participants for rating or labelling. The evaluation of AOIs or sentences is part of the same lab session, which typically lasts 45 to 90 minutes, depending on hardware calibration, the type and number of tasks to be performed, and the extend of user reported feedback that should be collected. 

\subsection{Technical setup}

For collecting and analysing eye-tracking data, we use the EyeLiveMetrics web browser plug-in \cite{Hienert2024}, which maps raw eye-tracking coordinates to words, text, images, and videos on web pages in naturalistic viewing or reading settings. While study participants perform the task, e.g., reading news online, the plug-in computes fixation-, saccade-, and reading metrics for fixated words and AOIs and saves them to a database. Via an API, eye-tracking metrics are retrieved by the EEG analysis component. 

For combining all input streams, Lab Streaming Layer (LSL) \cite{Kothe2024} is used. Raw gaze data from the eye tracker is streamed to Lab Streaming Layer (LSL). In our case, we use a customized version of the Tobii Pro Connector App\footnote{\url{https://github.com/labstreaminglayer/App-TobiiPro}} to forward all data points unaltered to LSL, including left and right pupil diameter and left and right gaze point validity. Similarly, EEG signals are also streamed into LSL. In our case, the OpenBCI GUI allows streaming EEG data to LSL. Keyboard and mouse input can be streamed to LSL using the LSL keyboard/mouse connector app\footnote{\url{https://github.com/labstreaminglayer/App-Input}. A continuous stream of x and y mouse pointer coordinates, and an event stream of mouse/keyboard interactions.}. All signal streams are synchronized and recorded using the LabRecorder app\footnote{\url{https://github.com/labstreaminglayer/App-LabRecorder}}, producing a unified XDF data file. Upon import using the pyxdf Python package\footnote{\url{https://github.com/xdf-modules/pyxdf}}, all modalities are temporally aligned based on the synchronized LSL clock\footnote{\href{https://labstreaminglayer.readthedocs.io/info/time_synchronization.html}{https://labstreaminglayer.readthedocs.io/info/time\_synchronization.html}}.

\subsection{Gaze data preprocessing}
Each gaze data sample typically contains binocular gaze point positions in display and in user coordinates, gaze origin coordinates, gaze validity flags, pupil diameters, and synchronized system and device timestamps.
LSL stream is, on one hand, merged with the other modalities and stored in the joint XDF file, and on the other hand, forwarded to EyeLiveMetrics for immediate online processing and persistent storage~\citep{Hienert2024}.

For each participant and stimulus (typically a webpage), EyeLiveMetrics computes a comprehensive set of per-word and per-sentence eye-movement metrics based on fixation and saccade data. Sentence-level metadata includes temporal boundaries of sentence fixation, sentence position and text, HTML identifiers, and basic length measures (word and character counts). Gaze measures comprise total and first-pass fixation durations and counts, time to first fixation, first fixation duration, regression-based measures (first-pass regressions, regression path duration, selective regression path duration), re-reading duration, look-back time, and the proportion of fixated words (see Table \ref{tab:eyetrackingmeasures}). To normalize for sentence length, fixation- and regression-based measures are additionally expressed per word and per character~\footnote{Except Rightward Saccades Average Length, Rightward Saccades Average Amplitude and Rightward Saccades Scan Speed}. Finally, saccadic behavior within each sentence AOI is quantified using counts, directions (rightward/progressive), lengths, amplitudes, scan speed, and corresponding word- and character-normalized variants. These features constitute the per-sentence output for all sentences fixated and read by a participant.

\subsection{EEG preprocessing}
\label{eeg_preprocessing}
In the current configuration, raw EEG data are processed using MNE-Python version 1.10.2 \cite{GramfortEtAl2013a} and typically include a notch filter to remove line noise and a band-pass filter (e.g., 0.5–15 Hz for frequency analysis or 0.1–30 Hz for ERP analysis). We use Independent Component Analysis (ICA) with the Picard algorithm~\cite{Ablin2017picard}~\footnote{\url{https://pypi.org/project/python-picard/}} to identify and remove artifacts such as blinks, eye movements, and muscle activity. Via the ICLabel plug-in integrated in MNE-Python, artifact-prone components are automatically labelled and excluded \cite{pion2019iclabel, li2022mne}\footnote{Filter settings, artifact correction strategy, thresholds, and component selection criteria are configurable and should be adapted to the specific hardware setup, signal quality, and analytical goals of a given study.}. 
\label{within-exp-eeg-analysis}
As gaze and EEG data are aligned with LSL, we extract the fixation onset for each fixated word in a sentence from the eye-tracking data to segment the EEG signal. We segment the cleaned EEG data into epochs relative to the onset of the first fixation of each word of a sentence. 
After segmentation, bad epochs are identified and excluded. Rejection criteria can be determined either dynamically~\cite{JasEtAl2017}
or a fixed maximum peak-to-peak threshold of e.g. 100 µV is applied. Segmented data are persisted for later in-depth analysis. 

\subsection{Real-time analysis and collection of user feedback}
Depending on the research questions, different analysis scripts can be plugged into the pipeline to select sentences or AOIs for user feedback. For example, metrics based on EEG band power as a proxy for cognitive load can be used to select sentences.
Similarly, FRP can be calculated for each word, as the Event-Related Potential (ERP) starting with the first fixation on the word, and used to select sentences for rating after the task. Likewise, any eye-tracking metric can be used to select sentences that participants should rate after the main task. Further, in the case of textual stimuli and reading at free choice, the read text itself can be analysed with e.g. the elfen package (Efficient Linguistic Feature Extraction for Natural Language Datasets) ~\cite{maurer2026elfen}, and the results can be used for grouping or selection of focused sentences or AOIs or for later analysis~\cite{Kraemer2026}. Figure \ref{fig:dataflow} summarizes the data collection flow. Gaze data is processed during acquisition, whereas EEG and other sensor data streamed via LSL are processed immediately after the task sessions.

\section{Measures}
Measures can be defined at the participant or at the stimulus level, e.g., for each sentence. 
Participant-level data, including socio-demographic profiles, general attitudes toward the research topic, and other relevant scales, are gathered using pre- and/or post-experiment questionnaires.
Depending on the research question, a set of questions or labelling tasks can be administered to participants at the stimulus level, e.g., for each AOI or sentence they have viewed that matches the selection criteria. Different selection criteria can be applied to select AOIs or sentences for user feedback. 
For each condition, a defined set of items (e.g., the top 10) will be presented to participants again for rating or labelling.
For within- and post-experiment analysis, word-level eye tracking metrics are aggregated by EyeLiveMetrics~\footnote{For a detailed description of each measure, see https://git.gesis.org/iir/eyelivemetrics} to the sentence level. 

For EEG, the system currently implements time-frequency analysis and can be extended with any MNE-Python analysis procedure, for example, FRP comparisons across conditions.

\begin{table}[htbp]
\centering
\footnotesize 

\caption{Eye-tracking measures. For all measures, both the summed values and their normalization based on the characters or words per sentence can be considered in the analysis.}

\setlength{\tabcolsep}{4pt} 

\begin{tabular}{ll}
\toprule
\textbf{Fixation Measures} & \textbf{Saccade Measures} \\ 
\toprule
Fixation Duration & Saccades Count: total number \\
Fixation Count & Rightward Count: number of rightward saccades \\
First Fixation Duration & Rightward Length: sum of lengths (px) \\
 & Rightward Avg Length: average length (px) \\
 & Rightward Avg Amplitude: average length (deg) \\
 & Rightward Scan Speed: speed (px/ms) \\
\toprule
\multicolumn{2}{l}{\textbf{Reading Measures}} \\ 
\toprule
\multicolumn{2}{l}{First Pass First Fix. Dur.: Sum of first fixation durations in first pass} \\
\multicolumn{2}{l}{First Pass Regression: Sum of regressions in the first pass} \\
\multicolumn{2}{l}{First Pass Duration: Sum of all fixation durations in first pass} \\
\multicolumn{2}{l}{Selective Reg. Path Dur.: Fixation durations before progressive fixation} \\
\multicolumn{2}{l}{Reg. Path Dur.: Durations including regressions before progressive fixation} \\
\multicolumn{2}{l}{Re-reading Duration: Regression Path Duration minus First Pass Duration} \\
\multicolumn{2}{l}{Look Back Time: Total fixation duration excluding first pass} \\ 
\bottomrule
\end{tabular}
    \label{tab:eyetrackingmeasures}
\end{table}

\label{us-gaze-measures}

\section{Experiment on Selective Exposure in News Search and Reading}
\label{sec:use_case}

The pipeline is currently used as an experimental environment to study selective exposure in online news reading. The broader research goal, addressed in a separate forthcoming publication, is to examine whether multimodal signals can be used to infer readers' agreement with the content they encounter. To this end, we implement a realistic news search and reading scenario in which participants search for, browse, select, and read online news articles.

\label{corpus}
The news corpus underlying the search scenario consisted of 192{,}685 German-language articles published between April~14,~2024 and April~14,~2025. 
To minimize visual confounds during reading, the application enforced a uniform presentation of textual content, following prior work~\cite{kosch2019investigating,KoschEtAl2020OneDoesNotSimplyRSVP}.
The news source was intentionally hidden to avoid influencing reading behavior through individual source preferences~\cite{knobloch2014choice}. While the system supports additional media consumption scenarios such as video or images, only text-based news reading was used in this study.
Participants are guided by the web-based application through the eight experimental steps illustrated in Figure~\ref{fig:expflow}. 

We focus on exploring the Theta band power activities, building on related work that has shown Theta oscillations to be correlated with memory retrieval and encoding~\cite{klimesch1994episodic,klimesch1997Theta,klimesch1999eeg} and with the processing of decision-consistent vs. inconsistent information~\cite{Fischer2013}. Because reading and processing online information might involve evaluating and resolving potentially conflicting information, Theta-band modulations could provide a theoretically grounded neural measure for studying how readers process congruent versus incongruent text.

\paragraph{Pre-experiment}
In the pre-experiment phase, we collected participants' general attitude towards two topics, climate change and migration policy, using a standardized scale~\cite{Mau2023}. 
We used an OpenBCI Cyton+Daisy board and an OpenBCI saline-based electrode cap at 125 Hz\footnote{\url{https://shop.openbci.com/products/all-in-one-gelfree-electrode-cap-bundle}} for EEG recording, with 16 channels
as well as additional reference and ground electrodes, and the OpenBCI GUI software for impedance monitoring and LSL streaming~\footnote{\url{https://github.com/OpenBCI/OpenBCI_GUI}}. Impedance was kept below $20 k\Omega$. The eye tracker (Tobii Pro Spectrum at 300 Hz) was calibrated using the Tobii Pro calibration tool (5 points). 

\paragraph{LSL Recording}

The LSL Recording phase began with a resting state EEG (1 minute with eyes open, 1 minute with eyes closed). Then, participants completed two news search tasks, one for each topic, migration policy or climate change. Topics were presented in randomized order. 

\paragraph{Post-Experiment}
\label{insight:influence-sel-criteria}
Participants then proceeded to give feedback. Because rating every fixated sentence would have been infeasible, participants were shown a manageable subset of sentences. For each topic, equal proportions of sentences were selected based on three predefined criteria:
(1) Longest fixation duration (sentences with the longest average fixation duration, normalized by sentence length in characters, reflecting increased visual attention during reading~\cite{just1980theory}), 
and (2) Theta power increase (sentences showing the strongest increase in theta-band power between 500\,ms and 1000\,ms or (3) between 1000\,ms and 1500\,ms after the onset of the first fixation of each word. Prior work associates theta increases in this time range with successful memory processes~\cite[e.g.][]{KLIMESCH2000}, and  motivated by reports of selective-exposure–related theta effects in this later period~\cite{Kraemer2025}). Using these criteria, a maximum of 60 sentences were rated per participant.


The system iterated over sentences selected according to these criteria and visually highlighted on the original web page. Participants were asked to indicate their familiarity with the sentence and assess their agreement with the sentence’s statement.
As a last step, participants were administered a questionnaire regarding their socio-demographic background and level of formal education. To assess personality traits, the German Big Five Inventory (BFI-10) was employed, as it provides a validated and efficient measure of the OCEAN dimensions in a short-scale format~\cite{Rammstedt2014}.

\section{Discussion \& Future work}

Accurate fixation detection is essential whenever downstream analyses depend on fixation-locked windows. The EyeLiveMetrics component for fixation detection and word-level AOI mapping has been validated against Tobii Pro Lab ground-truth, with Pearson correlations of $\leq 0.96$ across all fixation metrics \cite{Hienert2024}. Nonetheless, factors such as different eye tracking devices and sampling frequencies, calibration accuracy, display resolution, font size, and dynamic page content remain potential sources of noise that propagate to EEG epoching on fixation onsets and could thus bias FRP and time-frequency results \cite{Dimigen2011coregistration}.

Two directions for methodological extension follow from this. First, the impact of hardware capabilities, calibration quality, and text layout on gaze-to-word assignment should be systematically quantified by varying typographic and structural parameters under otherwise controlled conditions. Second, EEG preprocessing could be strengthened by integrating the eye-openness signal already available in the XDF stream, for instance, through ICA training~\cite{DIMIGEN2020opticat} or regression-based deconvolution approaches that explicitly model ocular events~\cite{EhingerDimigen2019unfold}.
This is particularly relevant for low-density mobile EEG setups, where ICA-based artifact correction is known to be less reliable than for high-density laboratory recordings.

Beyond these methodological refinements, the broader value of the proposed workflow lies in its modularity: components for stimulus presentation, signal acquisition, synchronization, and online analysis can be reused or replaced independently, allowing the workflow to be adapted to a wide range of research questions in naturalistic reading and related domains. We invite researchers interested in adopting or extending the pipeline to contact the authors, and we are committed to supporting its adaptation to new research contexts.

\section{Conclusion}

We introduced a fully integrated, real‑time multimodal pipeline that combines eye‑tracking, electroencephalography (EEG), and auxiliary sensor streams to investigate cognitive biases during naturalistic online reading. Through temporal synchronization and online gaze‑to‑text mapping, the system provides sentence‑ and word-level behavioural and neural metrics. 
The pipeline was applied in a use case of measuring the effect of selective exposure in online news reading, demonstrating how fixation‑ and EEG‑based selection criteria can be combined to isolate text passages that elicit heightened visual attention or deeper semantic processing.
The proposed pipeline thus offers an extensible instrument for testing, in naturalistic reading contexts, proxy relationships that cognitive science has largely established under tightly controlled experimental paradigms.


\section*{Acknowledgements}
This work is funded by the German Research Foundation (DFG) project ``Overcome Selective Exposure in Web Search by Considering Eye Movements and Physiological Signals''~(Project-ID~525041402). 
Furthermore, this work is supported by the German Research Foundation (DFG), CRC 1404: ``FONDA: Foundations of Workflows for Large-Scale Scientific Data Analysis''~(Project-ID~414984028), as well as by the German Federal Ministry of Research, Technology and Space (BMFTR): ``Claimguard''~(Project-ID~16KIS2483).
We thank \textit{DFG and BMFTR} for their valuable assistance.




\bibliography{references}
\bibliographystyle{elsarticle-harv}






\clearpage








\end{document}